\documentstyle[12pt]{article}
\begin{document}
\begin{flushright}
IJS-TP-96/16\\
1996\\
\end{flushright}
\renewcommand{\thefootnote}{\fnsymbol{footnote}}
\vspace*{7truecm}
\setlength{\baselineskip}{12pt}
\begin{center} 
THE ROLE OF RESONANCES IN RARE K DECAYS
\footnote{Talk presented at the Workshop on K-Physics, Orsay, France, 
30 May - 4 June, 1996}\\
\vspace{.5cm}
S. Fajfer\\
Institute ``J. Stefan'', Jamova 39, 1001 Ljubljana, Slovenia\\
\vspace{.5cm}
\end{center}
\centerline{ABSTRACT}
\vspace{0.5cm}
The decays $K \to  \pi l^{+} l^{-}$, 
$K \to  \pi \gamma \gamma $ 
and $K \to \pi \nu {\bar \nu}$ are investigated using the 
higher order terms of the chiral perturbation theory. 
The counterterms induced by strong, weak and electromagnetic 
interactions are determined assuming the resonance exchange. \\

\setlength {\baselineskip}{18pt}
\noindent
The chiral perturbation theory (CHPT) offers an 
useful framework to describe K decays$^{1-15)}$. 
The role of resonances in CHPT has been much better 
understood during the last few years$^{5-12)}$. 
We discuss the rare decays $K \to  \pi l^{+} l^{-}$, 
$K_L \to  \pi^0 \gamma \gamma $ and $K \to \pi \nu {\bar \nu}$. 
The first two decays obtain contributions of  
higher order terms in momentum of CHPT$^{1-4)}$. 
The $K \to \pi \nu {\bar \nu}$ decays are suppressed due to the 
GIM mechanism and the dominant contribution comes from the 
short distance dynamics. 
We analyse their long distance contributions. 
The $K^+ \to \pi^+ \nu {\bar \nu}$ decay amplitude 
obtains the leading contribution 
of $O(p^4)$, while the $K_{L,S} \to \pi^0 \nu {\bar \nu}$ amplitude
obtains the contribution of $O(p^2)$ chiral Lagrangians. \\
At the lowest order in momentum $O(p^2)$ the strong chiral Lagrangian is 
\begin{eqnarray}
{\cal L}_s^2 & = & \frac{f^2}{4} \{ Tr (D_{\mu} U^{\dag} D^{\mu} U\} 
+ Tr( \chi U^{\dag} + U {\chi}^{\dag})\},
\label{j1}
\end{eqnarray}
where $U = \frac{-i {\sqrt 2}}{f} \phi$, $f\simeq f_{\pi} = 0.093 $ GeV 
is the pion decay constant and
$\phi$ is a pseudoscalar meson matrix$^{1)}$. 
The covariant derivative is given by $D_{\mu} U = \partial_{\mu} U 
+ i U l_{\mu} - i r_{\mu} U$, $l_{\mu}$ and $r_{\mu}$ are external 
gauge field sources. The explicit chiral symmetry breaking induced 
by the electroweak currents of the standard model 
corresponds to the following choice:
\begin{eqnarray}
r_{\mu} & = & e Q [A_{\mu} - tan \theta_W Z_{\mu}], 
\label{j2}
\end{eqnarray}
\begin{eqnarray}
l_{\mu} & = & e Q [A_{\mu} - tan \theta_W Z_{\mu}] 
+ \frac{e}{sin \theta_W} Q_L^{(3)} Z_{\mu}
 +   \frac{e}{{\sqrt 2}sin \theta_W}[ Q_L^{(+)} W_{\mu}^{+}
+ Q_L^{(-)} W_{\mu}^{-}].
\label{j3}
\end{eqnarray}
Here $Q$'s are the electroweak matrices$^{1-4,12)}$. 
The matrix $\chi $ takes into account the explicit 
breaking due to the quark masses in the underlying QCD Lagrangian$^{13)}$. 
The strong chiral Lagrangian at $O(p^4)$ has 
$10$ phenomenological parameters which can 
be determined by low energy phenomenology$^{13)}$. 
The authors of ref.$^{6)}$
have considered 
the resonance contribution to the coupling constants of this  
${O}(p^{4})$ effective strong chiral Lagrangian. 
They have found that the resonance exchange can saturate the finite 
part of the conterterms. \\ 
In the weak sector at $O(p^4)$, by imposing 
$SU(3)_L \times SU(3)_R$ chiral symmetry, one introduces 
many undetermined couplings$^{7,14,15)}$. 
Due to the lack of experimental data additional assumptions 
about the weak Lagrangian are necessary in order to fix the unknown couplings. 
There are two procedures available: the factorization model 
and the weak deformation model$^{1-4,7)}$.
The factorization model $^{7-12)}$ relies on 
the charged weak current 
$J_{\mu} ={ \delta S_2}/{\delta l_{\mu}} + { \delta S_4}/{\delta l_{\mu}} + 
{ \delta S_6}/{\delta l_{\mu}} + \cdots$, 
where $S_i$ denotes the effective action of the order $p^i$. 
The effective weak Lagrangian is 
\begin{equation}
{\cal L}_w = 4 G_8 Tr ( \lambda_6 J_{\mu}  J^{\mu} ),
\label{w2}
\end{equation}
with $G_8 = {\sqrt \frac{1}{2}} G_F s_1 c_1 c_3 g_8$ defined in ref.$^{1)}$. 
From $K \to \pi \pi$ it was found that $|g_8|= 5.1$. 
Both models can be formulated without any reference to resonances. 
Since the renormalized part of the strong couplings 
in chiral Lagrangian at ${O}(p^{4})$ 
can be explained 
by resonance exchange$^{6)}$, 
it is reasonable to apply the same procedure to the weak Lagrangian at 
$O(p^4)^{7)}$.\\

$K\rightarrow \pi \gamma^{*}$ \\
It was shown$^{1-4)}$ that in the chiral perturbation theory at 
${O}(p^{2})$ $K\rightarrow \pi \gamma^{*}$ transitions 
are forbidden for a virtual photon
$\gamma^{*}(q)$ for any value of $q^{2}$.  
Combining the contributions coming from one-loop and counterterms,
induced by strong, weak and electromagnetic interactions$^{1-4)}$, 
the amplitudes for $K^{+} \rightarrow \pi^{+} \gamma^{*}$ 
and  $K_{S} \rightarrow \pi^{0} \gamma^{*}$  can be written as
\begin{eqnarray}
{\cal A}(K^{+} \rightarrow \pi^{+} \gamma^{*}) & = & 
\frac{G_{8} e}{(4 \pi)^{2}}
q^{2} (W_{+} + \Phi_{K} + \Phi_{\pi})(q^{2}) \epsilon^{\mu} (p^{\prime} 
+ p)_{\mu}, \label{e1}
\end{eqnarray}
\begin{eqnarray}
{\cal A}(K^{0}_{S} \rightarrow \pi^{0} \gamma^{*}) & = & 
\frac{G_{8} e}{(4 \pi)^{2}}
q^{2} (W_{S} + 2 \Phi_{K}) (q^{2}) \epsilon^{\mu} (p^{\prime} + p)_{\mu},
\label{e2}
\end{eqnarray}
where  $p$ and $p^\prime$ are pion's  and  kaon's momenta. 
The loop contributions $\Phi_{K}$ and $\Phi_{\pi}$ are 
determined in ref.$^{1-3)}$. The couplings 
$W_+ = - (16\pi^2/3) (W_1^r + 2 W_2^r -12 L_9^r) + 
(1/3) log(\mu^2/m_K m_{\pi})$ 
and $W_S = - (16\pi^2/3) (W_1^r- W_2^r) + (1/3)log(\mu^2/m_K^2)$,  
where $W_{1,2}^r$ have been defined in ref.$^{2-4)}$. 
Using the equations of motion for resonances, the authors of ref.$^{6)}$ 
have found that the finite parts of 
$L_{9}^{r}$, $L_{10}^{r}$ and $H_{1}^{r}$
get contributions from vector and axial-vector reconances:
\begin{eqnarray}
L_{9}^{V} & = & \frac{F_{V}G_{V}}{2 M_{V}^{2}}, \enspace
L_{10}^{V} = 2 H_{1}^{V} = - \frac{F_{V}^{2}}{4 M_{V}^{2}},  \enspace  
L_{9}^{A} = 0, \enspace 
L_{10}^{A} =  2 H_{1}^{A} = \frac{F_{A}^{2}}{4 M_{A}^{2}}, \enspace 
\label{e10}
\end{eqnarray}
where $M_{V}$ and $M_{A}$ are the octet masses of 
vector and axial-vector mesons.
The octet couplings $|F_{V}| = 0.154$ GeV $|G_{V}| = 0.069$ GeV are 
determined from 
the decay rates $\rho \rightarrow l^{+}l^{-}$ and $\rho \rightarrow 2 \pi$ 
respectively, while $F_{A} = 0.128$ GeV was determined in ref.$^{6)}$. 
We use $L_{9}^{r} = 6.9 \times 10^{-3}$,  
$L_{10}^{r} = -6.0 \times 10^{-3}$ and 
$H_{1}^{r} = 7.0\times 10^{-3}$ (set A) 
obtained in ref.$^{6)}$. 
We also use the numerical values $L_{9}^{r} = 7.0 \times 10^{-3}$, 
$L_{10}^{r} = -5.9 \times 10^{-3}$ and  
$H_{1}^{r} = -4.7 \times 10^{-3}$ (set B), and 
$L_{9}^{r} = 5.8 \times 10^{-3}$, 
$L_{10}^{r} = -5.1 \times 10^{-3}$, 
$H_{1}^{r} = -2.4 \times 10^{-3}$ (set C), 
calculated in the 
extended Nambu and Jona-Lasinio model$^{16)}$. 
Among three possible decays of $K \to \pi e^+ e^-$, only the decay rate of 
$K^{+} \rightarrow \pi^{+} e^{+} e^{-}$ 
has been measured. The branching ratio 
$BR(K^{+} \rightarrow \pi^{+} e^{+} e^{-}) = (2.99\pm 0.22)\times 10^{-7}$
from Brookhaven experiment gives the solution$^{17)}$ 
$W_{+} = 0.89_{-0.14}^{+0.24}$ from a fit to the high-$q^2$ 
spectrum.
The solution extracted from the measured decay 
rate in the same experiment corresponds to the value
$W_{+} = 1.2_{-0.5}^{+0.4}$. 
Using the weak deformation model$^{1-4,7)}$, it was found$^{6,10)}$ 
$W_1^r = 4(L_9^r + L_{10}^r  + 2 H_1^r)$ and $W_2^r= 4L_9^r$, leading to 
$W_+^{W,A} = -5.01$, $W_S^{W,A} = 4.50$, 
$W_+^{W,B} = 3.91$, $W_S^{W,B} = 3.49$, and 
$W_+^{W,C} = 2.80$, $W_S^{W,C} = 2.38$. 
Neither of them satisfies the experimental result. If the factorization 
model is used$^{7,8,10)}$, 
$W_1^r = 8(L_9^r + L_{10}^r  + 2 H_1^r)$ and $W_2^r= 8 L_9^r$, leading to 
$W_+^{F,A} = -4.87$, $W_S^{F,A} = 8.67$, 
$W_+^{F,B} = 2.69$, $W_S^{F,B} = 6.69$, and 
$W_+^{F,C} = 1.22$, $W_S^{F,C} 
= 4.46$. It means that the factorization approach, with the couplings C, 
can reproduce the experimental result. 
However, $W_{1,2}^W$, obtained with the help of  weak deformation approach, 
fulfills the condition for divergent parts of 
amplitudes$^{2,7)}$, what $W_{1,2}^F$ does not contain. \\
The decay  $K_{L} \rightarrow \pi^{0} e^{+} e^{-}$ is being investigated 
as a signal of direct $\Delta S = 1$ CP violation. In addition to a 
CP conserving 
process, which proceeds through two photon exchanges, there are two 
kinds of the CP violating decay$^{5)}$: 
one proportional 
to the well known parameter $\epsilon$  and the other 
direct CP violating effect. 
From our analyses$^{10)}$,  we calculate 
$BR (K_{L} \rightarrow \pi^{0} e^{+} e^{-})  =  1.15 \times 10^{-10}$,  
for $W_{S}^{F,C} = 4.46$, 
close to the experimental upper limit 
$4.3 \times 10^{-9}$, 
a result from$^{18)}$, 
and $5.5 \times 10^{-9}$ obtained by$^{19)}$. 
We support the idea$^{5)}$ that the existence of direct 
CP violating term in the branching 
ratio $BR (K_{L} \rightarrow \pi^{0} e^{+} e^{-})$  
cannot be seen without observing 
$BR (K_{S} \rightarrow \pi^{0} e^{+} e^{-})$.  \\

$K \to \pi \gamma \gamma$\\
There are two measurements of the branching ratio:    
$BR(K_L \to \pi^0 \gamma \gamma) = (1.70 \pm 0.3) \times 10^{-6}$ 
(NA31 result)$^{20)}$ and
$BR(K_L \to \pi^0 \gamma \gamma) = (1.86 \pm 0.60 \pm 0.60) \times 
10^{-6}$ (E731)$^{21)}$. 
The leading order contribution in CHPT ($O(p^4)$ order)$^{2)}$ 
comes from the loops. 
The rate is then $BR(K_L\to \pi^0 \gamma \gamma) = 0.67 \times 10^{-6}$. 
The corrections coming from physical intermediate states such as two pions 
were calculated ($O(p^6)$ order)$^{22)}$. 
In order to create a resonance exchange at $O(p^6)$, we follow the proposal of 
ref.$^{4)}$ and we 
write the strong Lagrangian keeping the terms leading in $\frac{1}{N_c}$ 
\begin{eqnarray}
{\cal L}_s^{R} & = & C_1^R F_{\mu \nu} F^{\mu \nu} 
Tr (Q^2 \partial_{\lambda} U^{\dag} \partial^{\lambda} U) 
+ C_2^R F_{\mu \alpha} F^{\mu \beta} 
Tr (Q^2 \partial^{\alpha} U^{\dag} \partial_{\beta} U).\label{2f1}
\end{eqnarray}
The $F_{\mu \nu}$ is the electromagnetic field strength tensor and the 
couplings $C_{1,2}^R$ are saturated by resonance exchange. 
We use the decay amplitudes for $V \to P \gamma$: 
\begin{eqnarray} 
{\cal L}(V \to P \gamma) = i e G_{VP \gamma} \epsilon_{\mu \nu \rho \sigma} 
F^{\mu \nu} Tr (V^{\rho}\{ \partial^{\sigma} U,Q\}).
\label{2f2}
\end{eqnarray}
Using the nonet assumption for vector mesons, the ideal mixing, 
and the data for $V \to P \gamma$, we calculate $|G_{VP \gamma}| 
= 7.7 \times 10^{-2}$. Eliminating vector mesons one easily derives
\begin{equation}
C_1^V = -\frac{1}{2} C_2^V = \frac{2 e^2G_{VP \gamma}}{M_V^2} 
= 3.92 \times 10^{-2} GeV^{-2}.
\label{2f3}
\end{equation}
We have shown$^{9,11)}$ that the contribution of scalar and tensor mesons is 
one order of magnitude 
smaller than this one. The amplitude for $K \to \pi \gamma \gamma$ can be 
decomposed into 
\begin{eqnarray}
{\cal A}(K(k) \to \pi(p) \gamma (q_1) \gamma (q_2))  & = & 
\epsilon_{\mu} (q_1)  \epsilon_{\nu} (q_2) [
\frac{A(y,z)}{m_K^2} (q_2^{\mu} q_1^{\nu}  -  
q_2\cdot q_1 g^{\mu \nu} ) \nonumber\\
+ 2 \frac{B (y,z)}{m_K^4}
(- k\cdot q_1 k \cdot q_2 k^{\mu} k^{\nu} &-& 
q_2\cdot q_1 k^{\mu} k^{\nu} + k \cdot q_1 q_2^{\mu} k^{\nu} 
+ k \cdot q_2 q_1^{\mu} k^{\nu})],
\label{2f4} 
\end{eqnarray}
where $A$ and $B$ are functions of the Dalitz 
variables $y= (k \cdot (q_1 - q_2) / m_K^2)$ 
and $z = (q_1 + q_2)^2 / m_K^2)$. 
In ref.$^{4)}$ the weak deformation approach was discussed for the calculation 
of the vector meson resonances. Using the factorization approach we derive
\begin{eqnarray}
A^R(y,z) & = & \frac{8 G_8 m_K^4 \alpha \pi}{9} (1 + g(\theta)) 
[4 C_1^R ( 1 - z + r_{\pi}^2 ) - C_2^R (1 + z - r_{\pi}^2)], 
\label{2f5}
\end{eqnarray}
\begin{eqnarray}
B^R(y,z) & = & \frac{16 G_8 m_K^4 \alpha \pi}{9} (1 + g(\theta)) C_2^R .
\label{2f6}
\end{eqnarray}
The function $g(\theta)= [m_K^2/(m_{\eta}^2 - m_K^2)]
(c - {\sqrt 2}s)(c +2 {\sqrt 2}s) 
+ [m_K^2/(m_{{\eta}^{\prime}}^2 - m_K^2)](s +{\sqrt 2}c)(s - 2 {\sqrt 2}c)$ 
($c\equiv cos \theta$, $s\equiv sin \theta$),  
is obtained by taking into account the mixing 
of the $\eta$ and $\eta^{\prime}$ as usual$^{11)}$. 
In the large $N_c$ limit $\theta \simeq -22^0$. 
Combining the resonance exchange with the contribution of 
loops at $O(p^4)$, the $O(p^6)$ unitarity corrections$^{22)}$, 
we calculate $BR(K_L \to \pi^0 \gamma \gamma) = 9.38 \times 10^{-7}$. 
Without resonances this braching ratio was found to be 
$BR(K_L \to \pi^0 \gamma \gamma) = 8.38 \times 10^{-7}$. 
The invariant mass spectrum of the two photons is rather well 
reproduced in our approach$^{11)}$.\\
The $K^+ \to \pi^+ \gamma \gamma$ decay obtain the $O(p^4)$ 
counterterms contribution $\hat{c} = 32\pi^2/3 [ 12 (L_9 + l_{10}) - W_1 - 
2W_2 - 2 W_4]$ defined in ref.$^{1-4)}$. The weak deformation model$^{4)}$ 
gives $\hat{c} = 0$, 
while the factorization model (set C) leads to $\hat{c} = -1.14$. 
The resonance exchange at $O(p^6)$ order gives the very small contribution 
to the amplitude 
\begin{equation}
A(y,z)^{(6)} = \frac{16 G_8 m_K^4 \alpha \pi}{9} 
C_1^R ( 1 - z + r_{\pi}^2 ).
\label{kplus}
\end{equation}
The branching ratio is found to be 
$BR(K^+ \to \pi^+ \gamma \gamma) = 4.19 \times 10^{-7}$.\\

$K \to \pi \nu {\bar \nu}$ \\
The $K \to \pi \nu {\bar \nu}$ decay amplitude is 
dominated by the short distance loop 
diagrams, due to the explicit dependence of the heavy quark mass. 
The $K^+ \to \pi^+ \nu {\bar \nu}$ decay amplitude can be written as 
\begin{eqnarray}
{\cal A} (K^+ \to \pi^+ \nu {\bar \nu}) & = & \frac{G_F}{{\sqrt 2}}
\frac{\alpha f_+}{2 \pi sin^2\theta_W} 
[V_{ts}^*  V_{td}\xi_t (m_t^2/M_W^2) 
 +  V_{cs}^* V_{cd} \xi_c (m_t^2/M_W^2)\nonumber\\ 
&+ &V_{us}^* V_{ud} \xi_{LD}] 
(k +p)^{\mu} {\bar u}(p_{\nu}) \gamma_{\mu} (1 - \gamma_5) v (p_{{\bar \nu}})
\label{knn1}
\end{eqnarray}
where $f_+$ is the form factor in ${\bar K}^0 \to \pi^+ e {\bar \nu}$ decay 
and $k$ and $p$ are K and $\pi$ meson's momenta respectively. 
The  decay amplitude $K^+ \to \pi^+ Z^0 \to \pi^+ \nu {\bar \nu}$ 
vanishes at $O(p^2)$$^{23-24)}$. 
The loop contributions were found to be roughly of order 
$10^{-7}$ smaller than that of short distance contributions$^{24)}$. 
We determine the long distance con
tribution $\xi_{LD}$ using the $O(p^4)$ chiral 
Lagrangian and assuming the factorization approach for the weak 
interactions 
\begin{eqnarray}
\xi_{LD} = 
\kappa 
\{8 m_K^2 L_5^r (2sin^2\theta_W  -1) 
 + q^2 [L_9^r (2 sin^2\theta_W  -1) + \frac{4}{3} L_{10}^r sin^2\theta_W  
+  H_1^r ( \frac{8}{3} sin^2 \theta_W -4)]\},
\label{knn2}
\end{eqnarray}
with $\kappa = {4 \pi^2 g_8}/{{\sqrt 2} M_Z^2 cos^2 \theta_W } $.
We calculate  the 
branching ratio 
$BR(K^+ \to \pi^+ \nu {\bar \nu})_{LD}^{A} = 0.17 \times 10^{-13}$, 
$BR(K^+ \to \pi^+ \nu {\bar \nu})_{LD}^{B} = 0.29 \times 10^{-13}$ and 
$BR(K^+ \to \pi^+ \nu {\bar \nu})_{LD}^{C} = 0.40 \times 10^{-13}$, using 
$L_9^r$, $L_{10}^r$ and $H_1^r$ from fits A, B and C. 
This is of order $10^{-3}$ 
smaller than the short distance contributions$^{25)}$.  
The decay 
$K_{S,L} \to \pi^0 \nu {\bar \nu}$ has the leading contribution 
of $O(p^2)$ in CHPT. However, this leads to 
the branching ratio 
$BR(K_{L} \to \pi^0 \nu {\bar \nu})_{LD} = 4.1 \times 10^{-18}$ 
and $BR(K_{S} \to \pi^0 \nu {\bar \nu})_{LD} = 1.4 \times 10^{-15}$, what is 
entirely negligable comparing the leading 
short distance contribution$^{25)}$. \\
\noindent
We can summarize that the 
resonance saturation of the counterterms and the use of 
factorization approach can  reproduce the 
experimental result for $W_+$ in $K^+ \to \pi^+ e^+ e^-$ decays, 
leading to the large value of $W_S$. It results in the prediction 
$BR(K_L \to \pi^0 e^+ e^-) 
= 1.15 \times 10^{-10}$. The resonance exchange, the loops of $O(p^4)$,  
and the unitarity corrections give $BR(K_L \to \pi^0 \gamma \gamma)= 
9.38 \times 10^{-7}$, when the factorization approach is applied. 
The long distance contributions 
in $K \to \pi \nu {\bar \nu}$ 
are much smaller than the leading short distance contributions. \\ 

\setlength {\baselineskip}
{12pt}

REFERENCES\\

\noindent 
1. G. Ecker, A. Pich and E. de Rafael, Phys. Lett. {B 189}
(1987) 363.\\
2. G. Ecker, A. Pich and E. de Rafael, Nucl. Phys. {B 291}
(1987) 692.\\
3. G. Ecker, A. Pich and E. de Rafael, Nucl. Phys. {B 303}
(1988) 665.\\
4. G. Ecker, A. Pich and E. de Rafael, Phys. Lett. {B 237}
(1990) 481. \\
5. J. F. Donoghue and F. Gabbiani, Phys. Rev {D 51} (1995) 2187. \\
6. G. Ecker, J. Gasser, A. Pich and E. de Rafael, 
Nucl. Phys. {B 321} (1989) 311.\\
7. G. Ecker, J. Kambor and D. Wyler, Nucl. Phys. {B 394} (1993) 101.\\
8. S. Fajfer, Z. Phys {C 61} (1994) 645.\\
9. S. Fajfer, Phys. Rev. {D 51} (1995) 1101.\\
10. S. Fajfer, Z. Phys. C 71 (1996) 307.\\
11. S. Fajfer, preprint TUM-31-89/95, IJS-TP-95/12, 
to appear in Nuovo Cim. A.\\
12. S. Fajfer, preprint HU-SEFT R 1996-05, IJS-TP-96/3.\\
13. J. Gasser and H. Leutwyler, Nucl. Phys. {B 321} (1985) 465, 517.\\
14. J. Kambor, J. Missimer and D. Wyler, Nucl. Phys. {346} (1990) 17.\\
15. J. Kambor, J. Missimer and D. Wyler, Phys. Lett. {261}
(1991) 496.\\
16. J. Bijnens, C. Bruno and E. de Rafael, Nucl. Phys. {B 390} (1993) 501.\\
17. C. Alliegro et al., Phys. Rev. Lett. {68} (1992) 278.\\
18. D. A. Harrris et al., Phys. Rev. Lett. {71} (1993) 3918.\\
19. K. E. Ohl et al., Phys. Rev. Lett. {64} (1990) 2755.\\
20. G. D. Barr et al., Phys. Lett. { B 242} (1990) 523, 
Phys. Lett. { B 284} (1992) 440.\\
21. V. Papadimitriou et al., Phys. Rev. D44, (1992) 573.\\
22. A. G. Cohen, G. Ecker, and A. Pich, Phys. Lett {B 304} (1993) 347.\\
23. M. Lu and M. B. Wise, Phys. Lett. {B 324} (1994) 461.\\
24. C. Q. Geng, I. J. Hsu and Y.C. Lin,  Phys. Lett. { B 355} (1995) 569.\\
25. G. Buchalla and A. J. Buras, Phys. Lett. B 333 (1994) 221.\\
\end{document}